\documentclass[journal=jacsat,manuscript=article]{achemso}

\usepackage[version=3]{mhchem} 

\author{Antoine Rolland, Lucien Pouget, Marc Brunel and Mehdi Alouini}
\email{mehdi.alouini@univ-rennes1.fr}
\alsoaffiliation[Second University]
{Institut de Physique de Rennes, D\'{e}partement d'Optique et Photonique, UMR CNRS-Universit\'{e} de Rennes 1 6251,\\ Campus de Beaulieu, Rennes cedex 35042, FRANCE\\}
\title[An \textsf{achemso} demo]
  {Optoelectronic down-conversion by four-wave mixing in a highly nonlinear fiber for millimeter-wave and THz phase-locking}

\begin{document}
\begin{abstract}
Optoelectronic down-conversion of a THz optical beatnote  to  a  RF   intermediate frequency is performed with  a standard Mach-Zehnder modulator followed by a zero dispersion-slope fiber.  The two interleaved optical spectra obtained by four-wave mixing are shown to contain more than 75 harmonics enabling to conveniently recover the phase noise of a beatnote at 770 GHz at around 500 MHz. This four-wave mixing down-conversion technique is  implemented in a two-frequency solid-state laser in order to directly phase-lock  its 168  GHz  beatnote to a 10 MHz local oscillator.   
\end{abstract}

\noindent Ultra-high resolution spectroscopy~\cite{tonouchi2007}, THz metrology~\cite{popovic2011}, or THz radio-astronomy~\cite{cliche2006,mayorga2012} are applications that need ultra-stable continuous-wave (cw) sources. While heterodyning two continuous optical waves in a photo-detector is a well-established approach to generate cw THz signals, stability usually relies on phase-locked loops (PLL)~\cite{williams1989,alouini2001,pillet2008,rolland2010}. When the heterodyne beat is at THz frequencies, the major issue is to down-convert the THz beat to RF frequencies in order to implement standard PLLs. In this perspective, an important experimental step up was made recently by using a nonlinear electro-optic modulator (EOM) to bridge the gap between widely spaced optical frequencies. The opto-electronic phase locked-loop principle (OEPLL) principle was successfully implemented to stabilize two-frequency laser beat signals at 100 GHz~\cite{rolland2011,pillet2012}, up to 250 GHz~\cite{rolland2012} in an improved set-up, and also to stabilize microcombs at 140 GHz mode spacing~\cite{delhaye2012}. In all cases, the frequency spacing is limited to the number of harmonics that the EOM can generate, typically 10.

Besides, four-wave mixing in optical fibers is an optical nonlinear effect which has been studied extensively for applications such as telecommunication when several optical carriers propagate simultaneously in the fiber~\cite{Agrawal2001}. In particular, the benefit of using this effect has been shown for the generation of optical frequency combs~\cite{sefler1998}. Indeed, a dispersion management in the fiber allows to generate very wide comb~\cite{fatome2006}. Moreover, the use of a zero-dispersion-slope highly nonlinear fiber offer an important potentiality to generate wide and flat optical frequency combs~\cite{cruz2008,Supradeepa2012,Yang2013,melo2013}. Thus, One can wonder if this nonlinear effect could be advantageously implemented to achieve an optoelectronic down-conversion from the THz to the RF band for the purpose of detecting an intermediate frequency which give access to the THz phase noise.

In  this  letter  we report  on a new solution including a highly non-linear fiber (HNLF)  to  bridge the wide frequency gap between the THz and GHz domains. Actually, by sending a two-frequency beam into the  HNLF, one expects to generate two interleaved frequency combs which should give access to an intermediate frequency signal carrying  the THz phase noise.

\begin{figure}[!h]
\centering
\includegraphics[width=12cm]{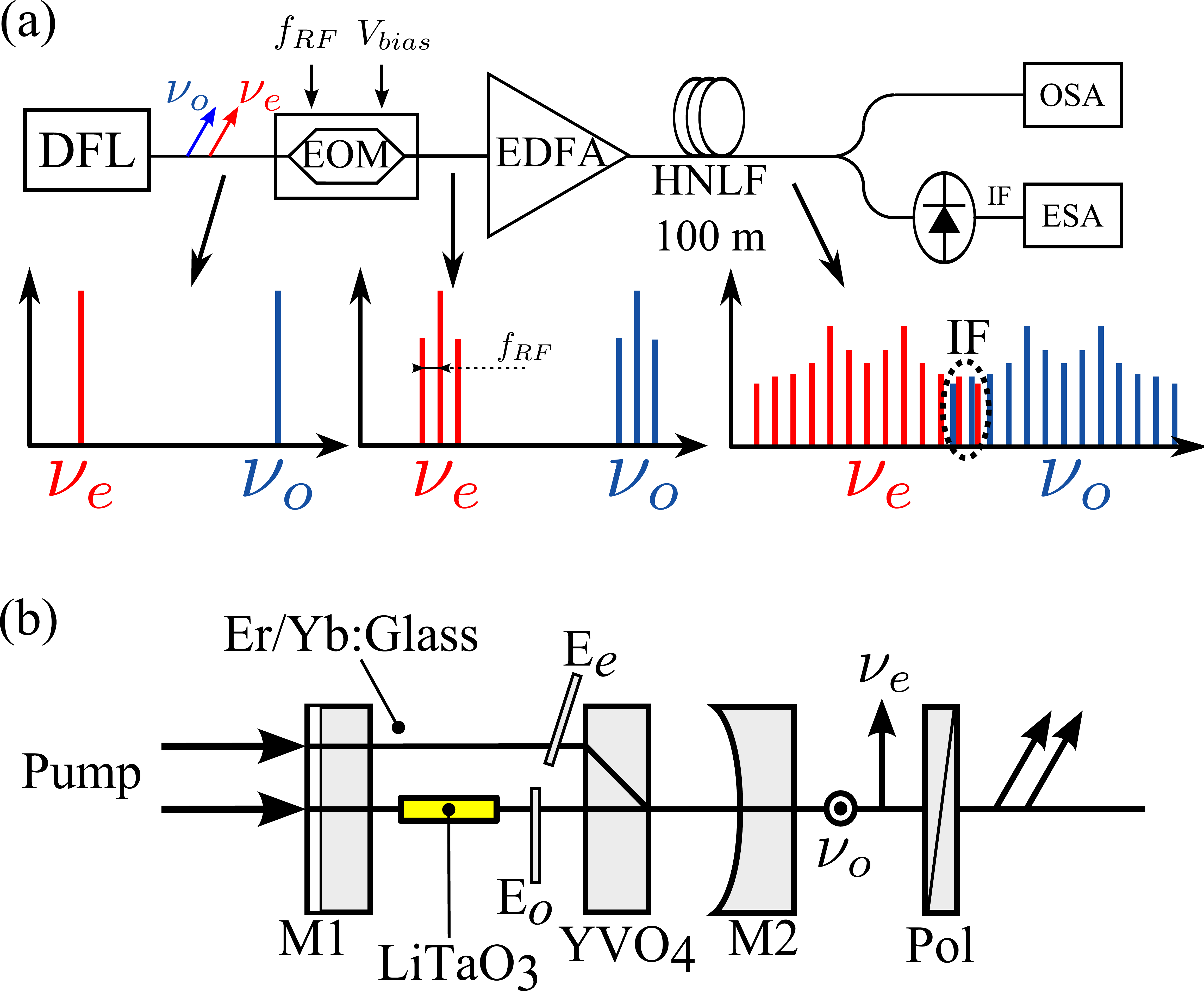}
\caption{(a) Optoelectronic down-conversion experimental setup. DFL, Dual-Frequency Laser; EOM, ElectroOptic Modulator; EDFA, Erbium Doped Fiber Amplifier; HNLF, Highly NonLinear Fiber; IF, Intermediate Frequency; OSA, Optical Spectrum Analyzer; ESA, Electrical Spectrum Analyzer. (b) Two-axis-propagation dual-frequency laser, see text for details.}
\label{figure1}
\end{figure}

The  principle of the experimental setup  is depicted  in  Fig.\ref{figure1}(a). It  involves two independent lasers or a single laser providing two optical frequencies, $\nu_e$ and  $\nu_o$, whose difference is in the THz range. The two beams are combined and then focused into  a  polarization-maintaining (PM)  fiber. Contrary to our previous work in which a dedicated nonlinear electrooptic modulator (EOM) was required~\cite{rolland2011}, we now use a standard EOM. This EOM is driven at $f_{RF}$ leading to the generation of two  sidebands  around  $\nu_e$ and  $\nu_o$. In order to optimize the four-wave mixing (FWM) efficiency in the HNLF, an erbium-doped fiber amplifier (EDFA) follows the EOM. The optical power sent into the 100 m-long HNLF  fiber can thus be adjusted from 20 dBm to 28 dBm. Note  that the  HNLF  dispersion slope  is chosen close to  0 ps/(nm$^{2}$.km),  which is required  to achieve optimal phase  matching of the generated frequency combs~\cite{cruz2008}. Finally, we adjust the bias voltage of the modulator, as well as the RF power in order to maximize the efficiency of the frequency comb generation.

In our experiment, the two optical frequencies are provided by a two-propagation-axis laser schematized in Fig.\ref{figure1}(b). The  active  medium  is a 1.5-mm-long phosphate glass doped Er/Yb. The  resonator  is closed by the input mirror M1 directly coated on the external face of the active medium and by  a 5-cm radius  of curvature mirror.  Due  to  an anti-reflection coated  10-mm-long YVO4   crystal  (cut  at 45$^\circ$  of its optical  axis)  inserted  into  the  cavity,  two  orthogonally  polarized  eigenmodes (labeled  respectively  o and e) are separated by 1 mm in the active medium while superimposed  at the output coupler. The active medium is pumped  at  980 nm  using  a laser  diode.  In  order  to efficiently pump  the  two eigenmodes, the  pump  beam  is split  into  two parallel  400-mW 100~$\mu$m-diameter beams separated by 1 mm.  To this  aim,  a second YVO$_4$ crystal (not shown in the figure) is inserted  between  the  pump  focusing lens and  the laser input  mirror.  To ensure  single mode oscillation  of each eigenpolarization and to adjust  independently their wavelengths,  we insert two  40 µm-thick  silica etalons  (E$_e$ and  E$_o$ ) coated  on both  sides for 30~\% reflection at  1550 nm.  Such a cavity  architecture leads to the  simultaneous oscillation  of two tunable  wavelengths which are linearly  cross-polarized.  By tilting  the etalons, we are able to tune independently the two wavelengths  over the  whole erbium  gain bandwidth. It leads to a beatnote adjustable between a few MHz and 2 THz~\cite{alouini1998,czarny2004}. Here  we keep  E$_o$   perpendicular to  the  ordinary   propagation  axis  while E$_e$ is tilted  to  sweep the  frequency difference  $\Delta\nu=\nu_e-\nu_o$. Furthermore, a LiTaO$_3$  electrooptic  crystal  is inserted  on the  ordinary  path  of the  cavity  offering a continuous  tunability of the beatnote.   This laser  is consequently turned into  a THz  voltage  controlled oscillator~\cite{rolland2011}. The  TEM$_{00}$  laser  output beam  is sent through a polarizer  oriented  at  45$^\circ$  in order to make the two modes beating before entering the EOM.

\begin{figure}[!h]
\centering
\includegraphics[width=12cm]{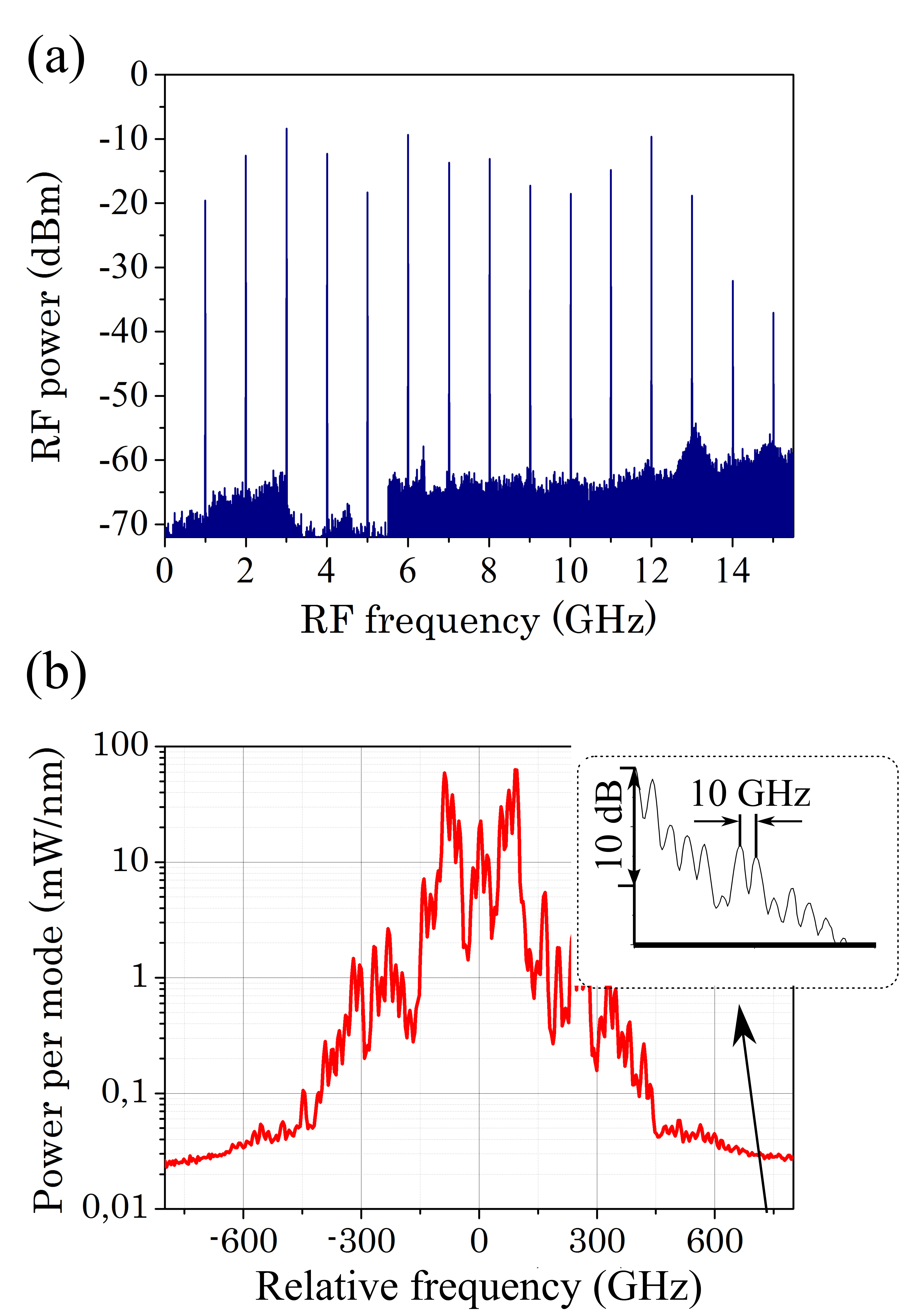}
\caption{(a) Electrical spectrum, with $f_{RF}$ = 1 GHz. (b) Optical spectrum (RBW 0.05 nm). The inset shows the optical spectrum beyond 700 GHz from the carrier.}
\label{figure2}
\end{figure}

Experimental results  are presented in Fig.~\ref{figure2}. In order  to  evaluate  the  electrical  spectrum of the generated combs, we first send through the EOM one optical frequency and we turn  $f_{RF}$ to a low frequency (1~GHz). Fig.~\ref{figure2}(a)  displays the  15 first  harmonics  of $f_{RF}$ observed with a 16~GHz bandwidth photodiode after propagation along the HNLF. The frequency comb has a  quite  flat  amplitude,  which  greatly  improves  previous  achievements  using a nonlinear EOM or a phase modulator~\cite{rolland2011,pillet2012}. Although $f_{RF}$ is low, the entire comb cannot be displayed because of the limited cut-off frequency of our photodiode. To observe higher harmonics and then evaluate the comb span, we now set the frequency  modulation $f_{RF}$  at 10 GHz and monitor the comb in the optical domain.  The generated comb spans 1 THz leading to measurable lines up the 75th harmonics, as shown on the optical spectrum of Fig.~\ref{figure2}(b) .

\begin{figure}[!h]
\centering
\includegraphics[width=12cm]{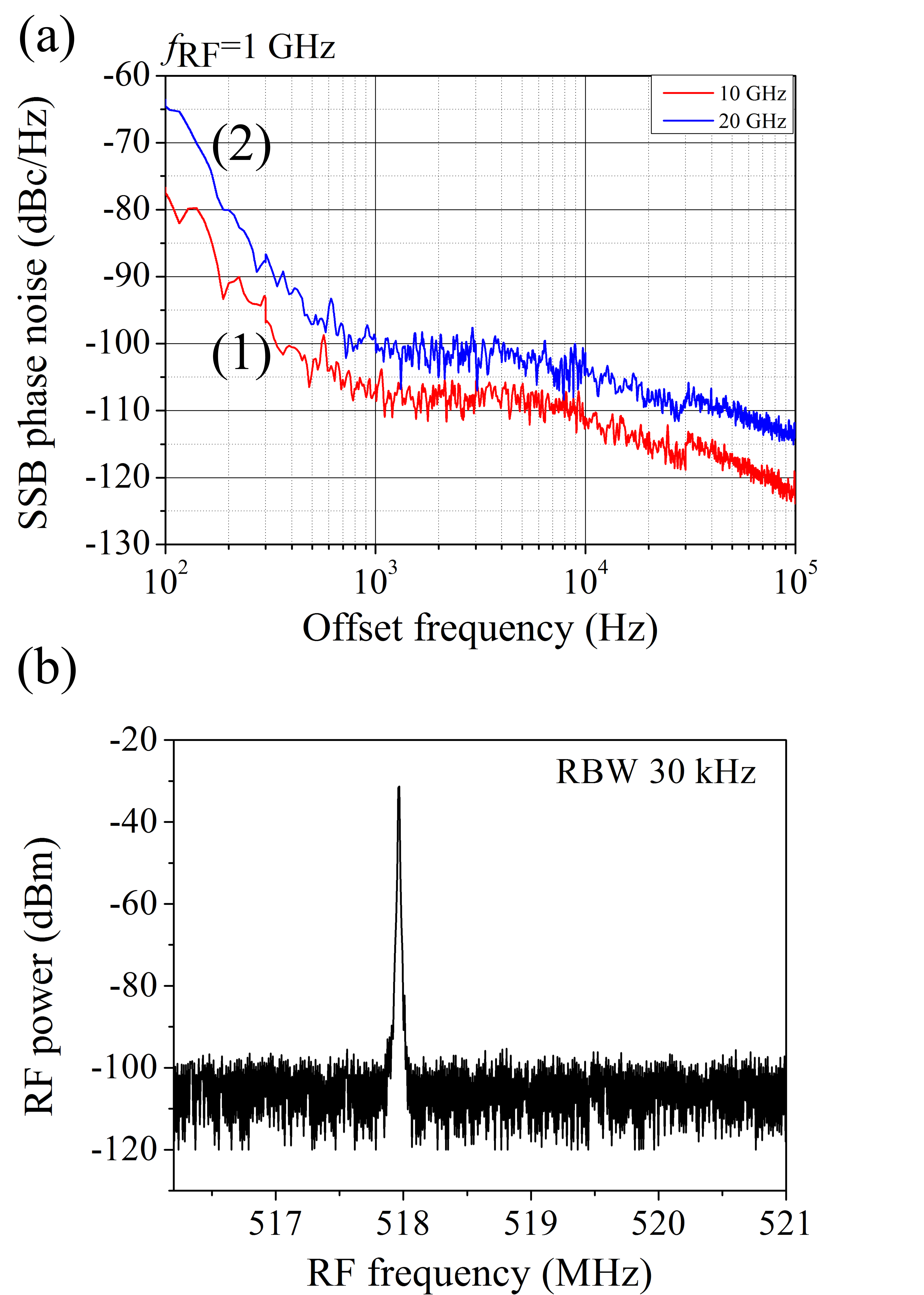}
\caption{(a) SSB Phase noise spectra. (1) N = 10 (10GHz) and (2) N = 20 (20GHz). (b) Intermediate frequency for a frequency difference $\Delta\nu=\nu_e-\nu_o=$770 GHz (RBW 30 kHz).}
\label{figure3}
\end{figure}

An accurate spectral analysis of the comb lines is performed by measuring the RF phase noise of the different harmonics numbered N.  To this aim, we turn again to a modulation frequency $f_{RF}$ of 1 GHz. For illustration purpose, two phase noise spectra  for (1)  $N$ = 10 (10 GHz)  and  (2)  $N$ = 20 (20 GHz) are reported in  Fig.~\ref{figure3}(a). The expected phase noise degradation of $20\log(N)$~\cite{drouin2005} is confirmed experimentally. One can thus conclude that the 100m-long fiber does not bring additional noise degradation at least for the first 20th harmonics generated by FWM. Let us remind here that our goal is to down convert a THz beatnote in the GHz range. We now send to the modulator the two optical frequencies at the same time. Setting the frequency difference of the laser to $\Delta\nu=\nu_e-\nu_o=$770~GHz, we were able to measure  an intermediate frequency  between  the interleaved combs  at  around  500 MHz  (see  Fig.~\ref{figure3}(b)). A dynamic  range  as high as 70~dB is then obtained  on this intermediate frequency with a 30~kHz measurement bandwidth. This high dynamic range makes it possible to evaluate the spectral purity  of the THz beatnote without any electrical component operating  in the  THz range.  Indeed,  the highest cut-off frequency in the  experiment  is that of the EOM, i.e., 10~GHz. Although demonstrated with a dual frequency laser, this FWM-assisted down-conversion technique is obviously well suited for two independent solid-state or semiconductor lasers. Moreover, the technique demonstrated here involves only widely spread and commercially available components in contrast to ref~\cite{rolland2011}.

\begin{figure}[!h]
\centering
\includegraphics[width=12cm]{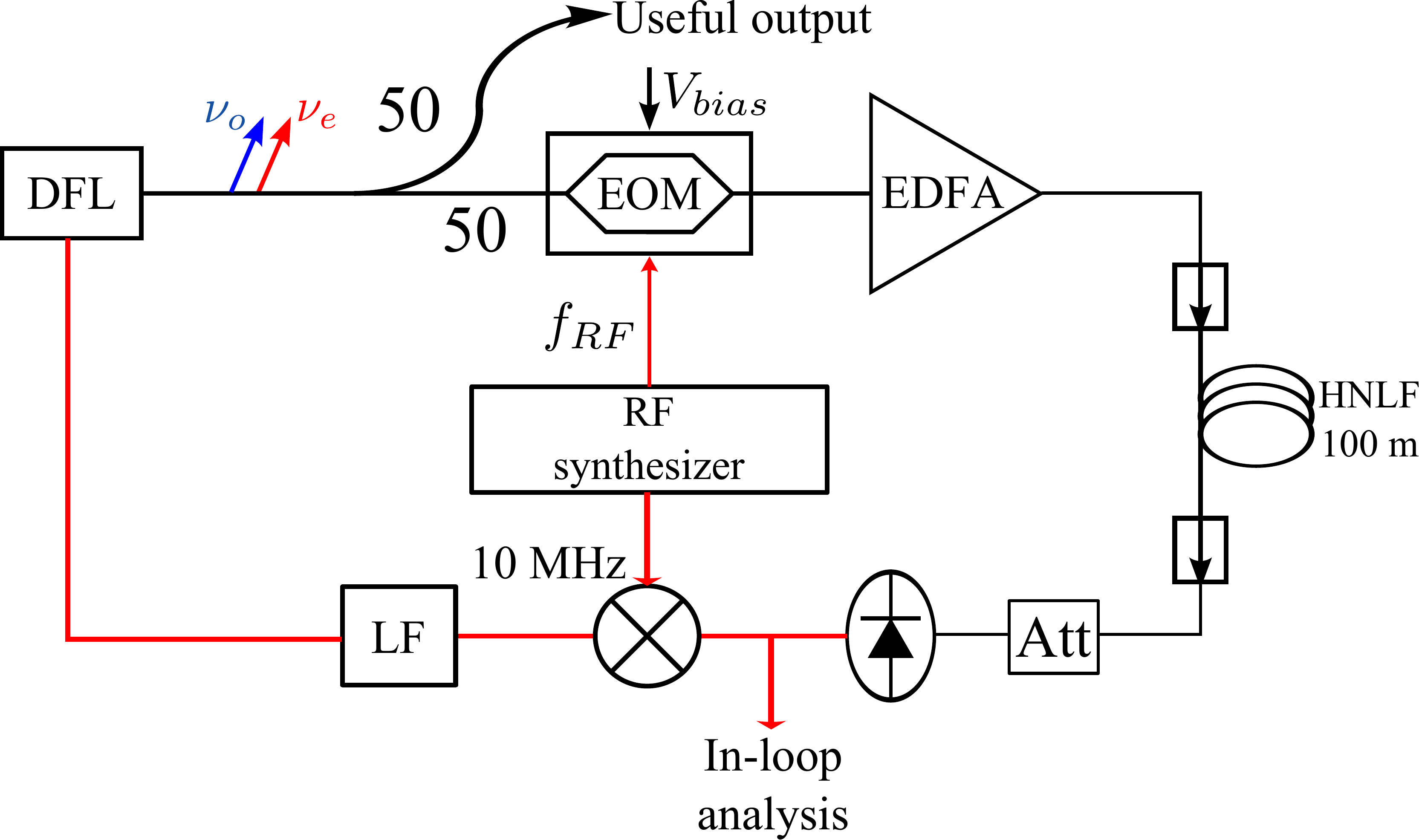}
\caption{Complete experimental setup for the FWM-based OEPLL. Att, optical attenuator; LF, Loop Filter. }
\label{figure4}
\end{figure}

\begin{figure}[!h]
\centering
\includegraphics[width=12cm]{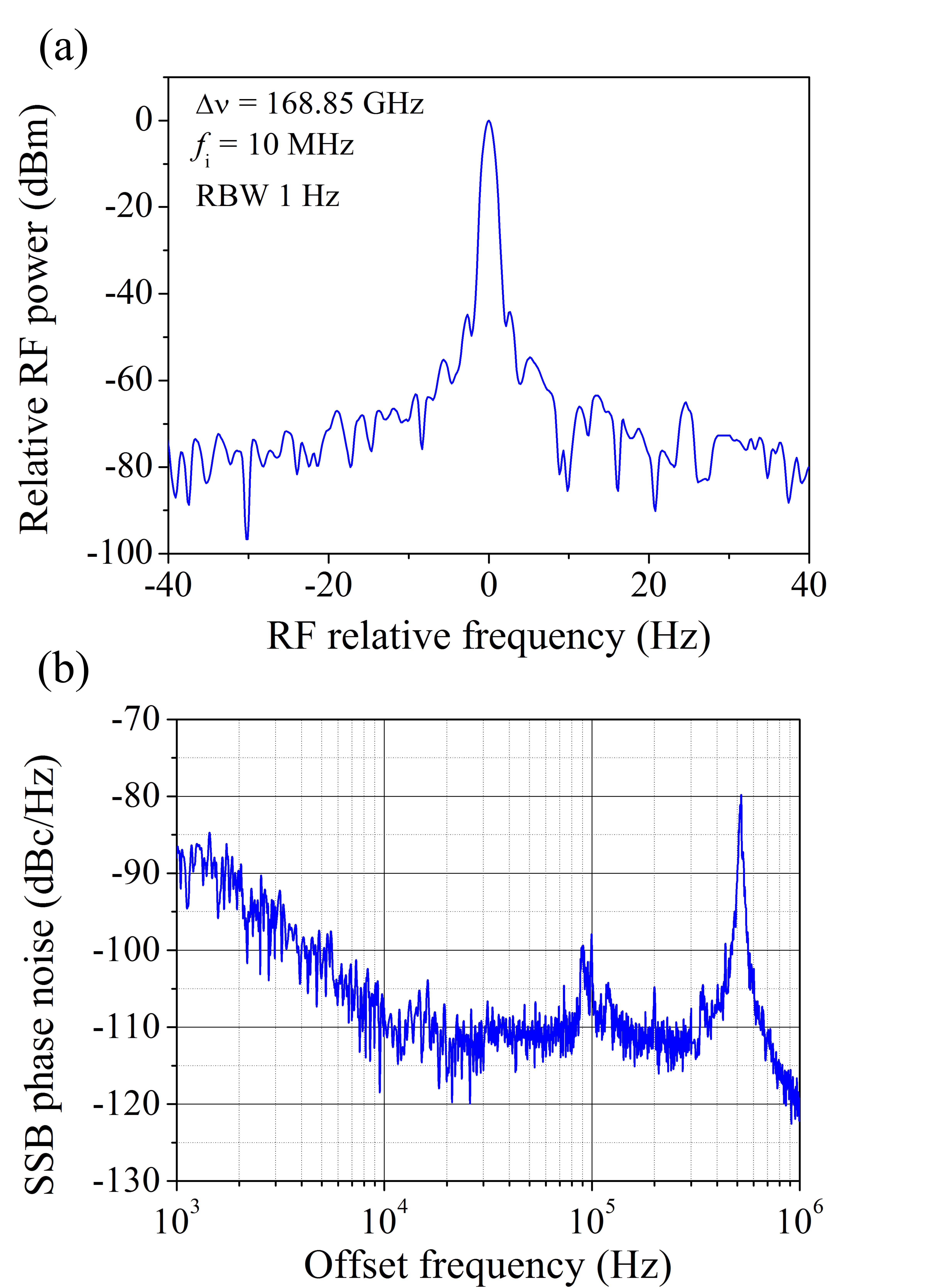}
\caption{(a)Electrical spectrum of the stabilized intermediate frequency $f_i$ for a frequency difference $\Delta\nu=$168.85~GHz with a 1~Hz resolution bandwidth and a 80~Hz span. (b) Absolute SSB phase noise of the intermediate frequency}
\label{figure5}
\end{figure}

Now, one can wonder if we can use this approach in order to actively stabilize the optical beatnote $\Delta\nu$ through the generated intermediate frequency. To this aim, we implement the experimental servo-loop depicted  in Fig.~\ref{figure4}. For the purpose of phase locking , it is convenient to down convert the laser beatnote directly to $f_i$ = 10 MHz, i.e., at the frequency of the quartz oscillator that will be used as frequency reference. This is obtained by adjusting precisely the modulation frequency $f_{RF}$. We then electrically mix $f_i$ with  the  10 MHz reference quartz  oscillator  of the  RF synthesizer leading to a  DC voltage  proportional to the  phase  error.  Through a loop filter (100 kHz bandwidth),  we finally apply  the  error  signal on the  LiTaO$_3$  crystal located inside the laser cavity. The  laser frequency  difference $\Delta\nu$ is set at around 170~GHz. We have chosen this value for OEPLL demonstration purpose, instead of 770~GHz as before, because our laser was actually optimized in terms of optical power and long-term stability for optimal operation around $\Delta\nu=$100~GHz.  To make the intermediate frequency $f_i$ close to 10~MHz, the EOM frequency $f_{RF}$ has to be adjusted to 10.5525~GHz. The  value of the frequency difference being given by $\Delta\nu=f_i+2Nf_{RF}$, where N is the  locking harmonic order, its exact value is actually  $\Delta\nu=$168.85~GHz with $N=$ 8. By closing the loop we stabilize the intermediate frequency and, consequently, the frequency  difference $\Delta\nu$. This  result  is reported in Fig.~\ref{figure5}. Fig.~\ref{figure5}(a) displays the electrical spectrum of the intermediate frequency $f_i$ when the loop is closed.  The full width  at half maximum is measured    to be lower than  1~Hz  limited  by the resolution of our electrical  spectrum analyzer.  This leads to an in-loop relative instability $\delta\nu$/$\Delta\nu$ at the 10$^{-11}$  level. As already mentioned, the signal-to-noise ratio  is greatly  improved  as compared  to the  results  presented in Ref.~\cite{rolland2011}, 80~dB here. It is worthwhile  to notice that the spectral  purity  of the stabilized   beatnote is governed  by  the  spectral  purity of the  RF  synthesizer  delivering  $f_{RF}$  and  the  10~MHz reference.  To evaluate precisely the  noise closer to the  carrier,  we measure  now the  single sideband  (SSB)  phase  noise of the  intermediate frequency  (see Fig.~\ref{figure5}(b)). The absolute phase noise is found to be $-$110~dBc/Hz at 10 kHz from the carrier.  However, one can notice a resonant noise at 550 kHz whose origin is still not identified. Indeed, this frequency  corresponds neither to the  relaxation oscillation frequency  of the  laser  (90 kHz) nor to the free spectral range of the optelectronic loop which is estimated at around 180 kHz. Further studies are actually devoted to figure out the origin of this excess noise.

In conclusion, we propose and demonstrate an optoelectronic down-conversion of a THz beat note to RF frequencies using four-wave mixing in a highly nonlinear  fiber. This principle is illustrated using a dual frequency laser whose frequency difference is set at 770~GHz.  Two interleaved combs are then generated around each optical carrier using a standard Mach-Zehnder modulator at 10~GHz followed by a highly nonlinear zero dispersion slope fiber. These combs are shown to contain more than 75 harmonics offering an intermediate frequency in the MHz range with an SNR of 70~dB. Moreover we show that the four wave mixing mechanism does not bring any additional degradation of the measured phase noise, at least up to the 20th harmonic. Hence, the intermediate frequency gives access to the phase noise of the THz beatnote using only commercially available optoelectronic components operating at room temperature. Although demonstrated using a dual frequency laser, the proposed down conversion technique could apply to a couple of detuned single mode lasers including semiconductor lasers. 

In a second part, this four-wave mixing down-conversion technique is advantageously implemented in an optoelectronic phase locked loop in order to directly phase-lock the laser beat note at 168~GHz  to a 10~MHz local oscillator. The beat note linewidth is then reduced down to 1 Hz corresponding to an in-loop relative  instability  of 10$^{-11}$ . In these conditions, the  phase  noise level is measured to be $-$110~dBc/Hz at 10~kHz from the carrier and is shown to be limited by the RF synthesizer phase noise. This  second part of the work is a new step  for the  generation  of ultra- high spectral  purity  THz waves. Further studies include a detailed analysis and understanding of the resonant phase noise we observed at 500~kHz from the carrier and which is still unexplained. Moreover, out-of-loop measurements will be undertaken  in order to fully characterize the stability of the THz beatnote. According to the significantly high down-conversion SNR offered by this four wave mixing approach, the phase locking of two independent lasers is now envisaged using a loop filter with increased bandwidth.

\bigskip

The authors are very grateful to Goulc'hen Loas, Ludovic Frein, Cyril Hamel, Steve Bouhier and Anthony Carr\'{e} for their help. This work is partially funded by R\'{e}gion Bretagne, Rennes M\'{e}tropole, FEDER, and DGA.

\bibliography{ref_FWM}


\end{document}